\documentclass{egpubl}
\usepackage{eg2018}

 \ShortPresentation      
 \electronicVersion 

\ifpdf \usepackage[pdftex]{graphicx} \pdfcompresslevel=9
\else \usepackage[dvips]{graphicx} \fi

\PrintedOrElectronic

\usepackage{t1enc,dfadobe}

\usepackage{egweblnk}
\usepackage{cite}

\usepackage{amsmath,amsfonts,amssymb}
\usepackage{algorithm}
\usepackage[noend]{algpseudocode}
\usepackage[inline]{enumitem}
\usepackage{verbatim}
\usepackage{gensymb}
\usepackage{textcomp}
\usepackage[hypcap=true]{caption}
\usepackage[justification=centering]{caption}
\usepackage{subcaption}
\title[Accurate control of a pan-tilt system based on parameterization of rotational motion]%
      {Accurate control of a pan-tilt system based on parameterization of rotational motion
}

\author[Byun, JungHyun et al.]
{\parbox{\textwidth}
	{\centering 
		JungHyun Byun, SeungHo Chae, and TackDon Han
	}
        \\
{\parbox{\textwidth}
	{\centering 
		Media System Lab, Department of Computer Science, Yonsei University, Republic of Korea
       } 
}
}

\begin{document}


\maketitle
\begin{abstract}
A pan-tilt camera system has been adopted by a variety of fields since it can cover a wide range of region compared to a single fixated camera setup.
Yet many studies rely on factory-assembled and calibrated platforms and assume an ideal rotation where rotation axes are perfectly aligned with the optical axis of the local camera.
However, in a user-created setup where a pan-tilting mechanism is arbitrarily assembled, the kinematic configurations may be inaccurate or unknown, violating ideal rotation.
These discrepancies in the model with the real physics result in erroneous servo manipulation of the pan-tilting system.	
\\
In this paper, we propose an accurate control mechanism for arbitrarily-assembled pan-tilt camera systems.
The proposed method formulates pan-tilt rotations as motion along great circle trajectories and calibrates its model parameters, such as positions and vectors of rotation axes, in 3D space.
Then, one can accurately servo pan-tilt rotations with pose estimation from inverse kinematics of their transformation.
The comparative experiment demonstrates out-performance of the proposed method, 
in terms of accurately localizing target points in world coordinates, after being rotated from their captured camera frames.


\end{abstract}  
\section{Introduction}

A pan-tilt platform consists of two motors, each rotating in \emph{pan} and \emph{tilt} directions. These 2 degrees of rotation freedoms grant the mounted system theoretically 360 degrees field-of-view.
This is particularly beneficial to computer vision applications, for cameras can only obtain data from a field of view that is directed by the optical axis at a time \cite{niu2017calibration}. To obtain scene information from a larger field of view, the camera had to be translated or rotated to capture a series of images. 
Various applications utilizes pan-tilt platforms from obvious video surveillance \cite{davis2003calibrating} to video conferencing, human-computer interaction, and augmented/mixed reality \cite{wilson2012steerable, byun2017air}.

Despite continuous endeavors of the literature on the pan-tilting mechanics, many studies based their ground on fully factory-assembled and calibrated platforms such as in \cite{davis2003calibrating, wilson2012steerable, li2015method}, which price from several hundreds to thousands of dollars. This may be problematic because the pan-tilting model may work seemingly error-free even if assuming an ideal rotation where rotation axes are aligned with the camera optical axis perfectly orthogonally, due to sound reliable assembly quality.

On the other hand, there is growing interest in user-created robots \cite{park2011philosophy}, where users create their own version of robots with such as consumer kits on the market \cite{byun2017air}. In this case, it is unlikely that the exact kinematic specifications will be provided, for example in a CAD file format. Moreover, the mechanism may be fabricated and assembled in an do-it-yourself or arbitrary manner. 
All these factors attribute to erroneous, if even existent, kinematic equations, rendering 
kinematic control methods obsolete.

In this paper, we would like to address the issue of accurate pan-tilting manipulation even the pan-tilting kinematics are loosely coupled, or often unavailable.
More specifically, we propose an operating mechanism of an arbitrarily assembled pan-tilt model with loose kinematics based on rotational motion modeling of the mounted camera.
Our method is based on the pan-tilt model that is general enough to calibrate and compensate for assembly mismatches, such as skewed rotation axes or off-origin rotations.

\section{Related Work}

Calibrating pan-tilt motion of the camera has been broadly studied in the computer vision field, especially surveillance using pan-tilt-zoom (PTZ) camera. For example,
Davis and Chen in \cite{davis2003calibrating} presented a method for calibrating pan-tilt cameras that incorporates a more accurate complete model of camera motion. Pan and tilt rotations were modeled as occurring around detached arbitrary axes in space, without the assumption of rotation axes aligned to camera optical axis.
Wu and Radke in \cite{wu2013keeping} introduced a camera calibration model for a pan-tilt-zoom camera, which explicitly reflects how focal length and lens distortion vary as a function of zoom scale. Using a nonlinear optimization, authors were able to accurately calibrate multiple parameters in one whole step. They also investigated and analyzed multiple cases of pan-tilt errors to maintain the calibrated state even after extended
continuous operations.
Li et al. in \cite{li2015method}  presented a novel method to online calibrate the rotation angles of a pan-tilt camera by using only one control point. By converting the non-linear pan-tilt camera model  into a linear model according to sine and cosine of pan and tilt parameters, a closed-form solution could be derived by solving a quadratic equation of their tangents.

In the optics and measurement literatures, studies regarding calibration methods for a turntable or rotational axis have been proposed. 
Chen et al. in \cite{chen2014rotation} fixed a checkerboard plate on a turntable and captured multiple views of one pattern. Retrieved 3D corner points in a 360\textdegree\ view were used to form multiple circular trajectory planes, the equation and parameters of which were acquired using constrained global optimization method.
Niu et al. in \cite{niu2017calibration} proposed a method for calibrating the relative orientation of a camera fixed on a rotation axis, where the camera cannot directly `see' the rotation axis. Utilizing two checkerboards, one for the rotating camera and one for the external observing camera, they were able to calibrate the relative orientation of the two cameras and rotation axis represented in the same coordinate system.

In this paper, we propose 
an operating mechanism of an arbitrarily assembled pan-tilt model with loose kinematics
based on rotational motion modeling of the mounted camera.
Our contributions can be summarized as follow:

\begin{itemize}
\item First, the proposed method models and calibrates the rotation of a generic pan-tilt platform by recovering directions and positions of its axes in 3D space, utilizing an RGB-D camera.
\item Second, the proposed method is capable of manipulating servo rotations with respect to the camera, based on the inverse kinematic interpretation of its pan-tilt transformation model.
\end{itemize}

\section{Pan-Tilt Rotation Modeling}
\begin{figure}[!htbp]
\centering

\begin{subfigure}[b]{0.43\linewidth}
\centering
\includegraphics[width=0.99\linewidth]{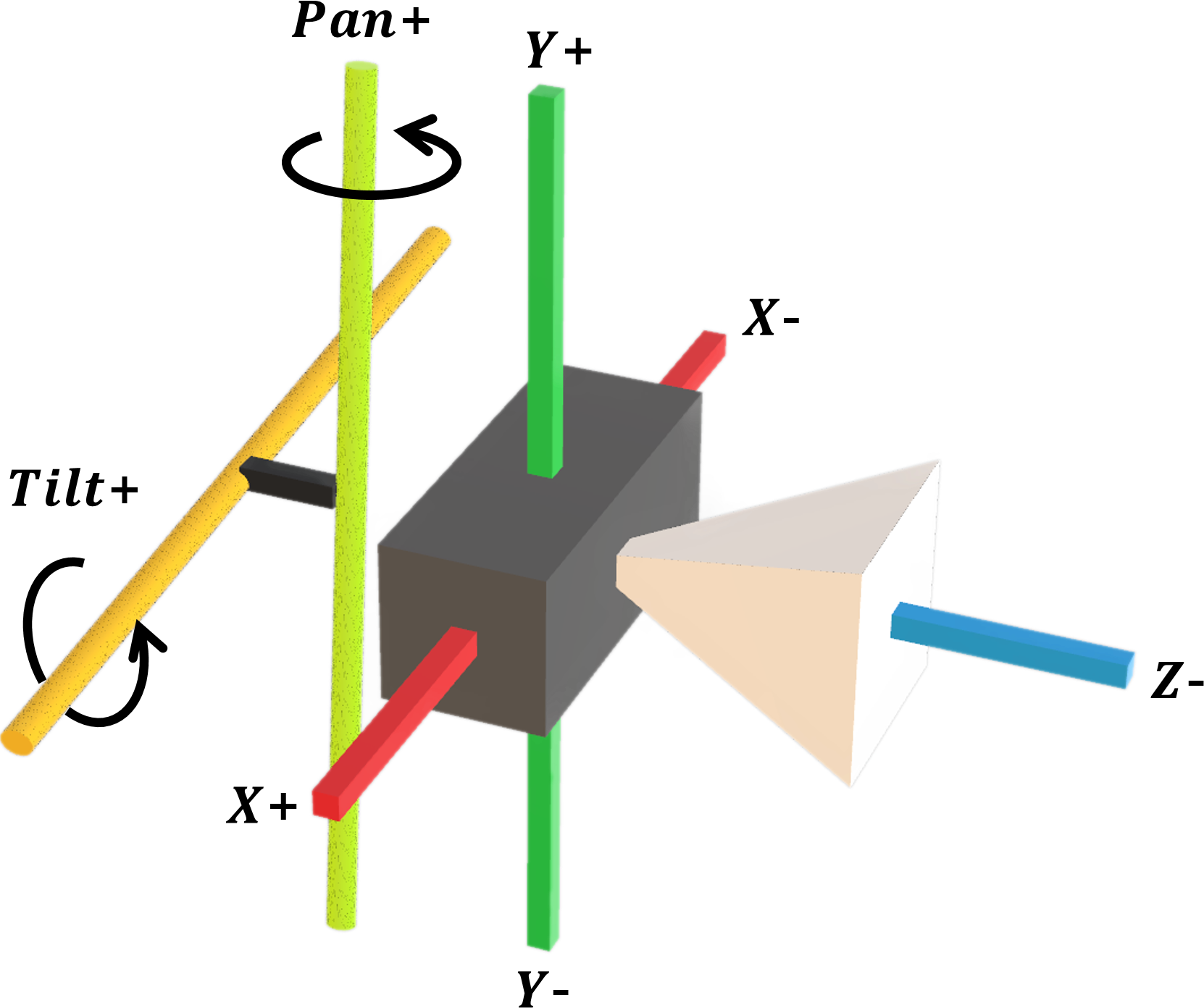}
\caption{\label{fig:pan_tilt_model}
Pan-tilt assembly model.}
\end{subfigure}%
~
\begin{subfigure}[b]{0.57\linewidth}
\centering
\includegraphics[width=0.99\linewidth]{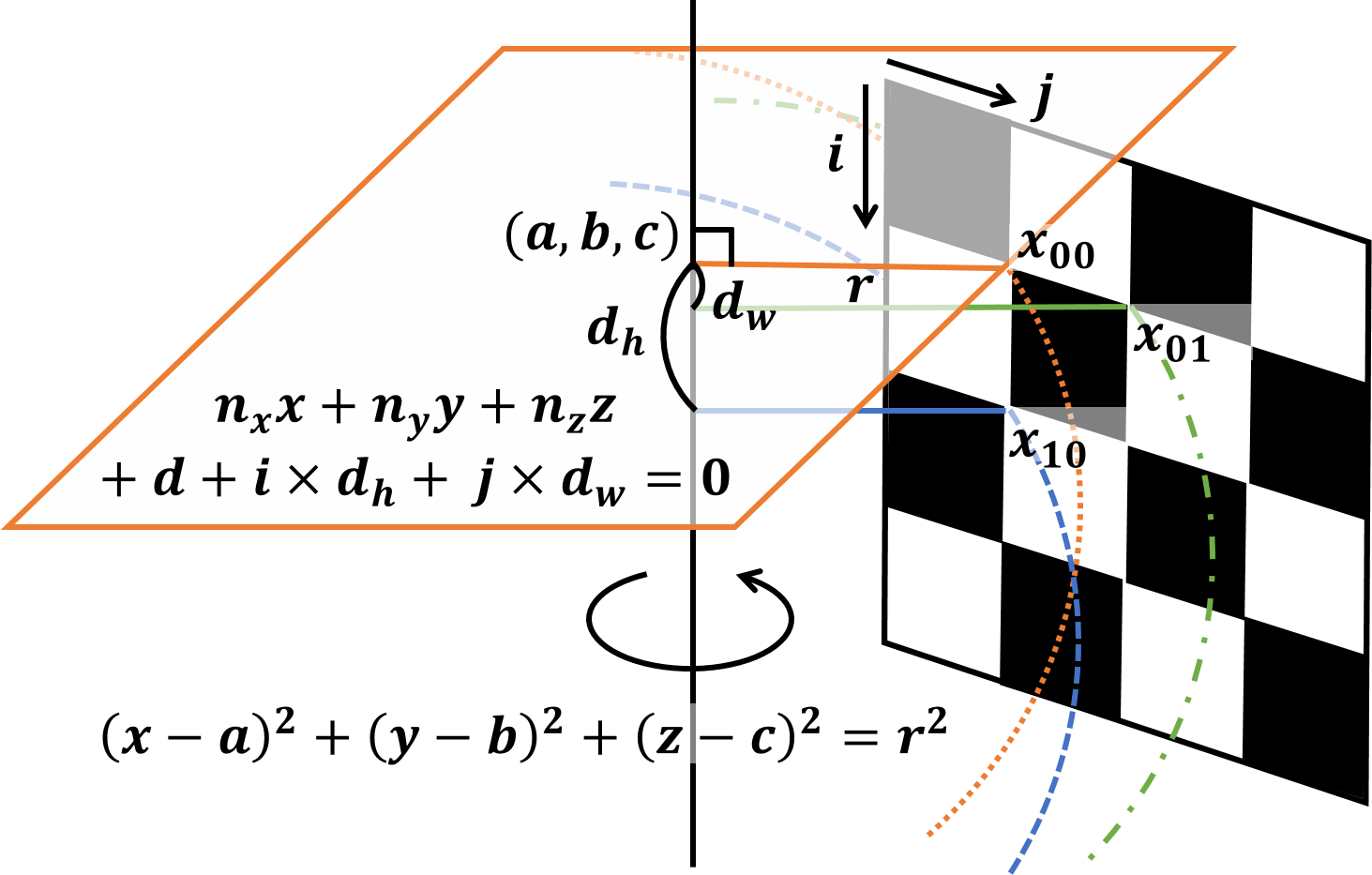}
\caption{\label{fig:rotation_formation}
Rotation model formation.}
\end{subfigure}

\caption{\label{fig:pan_tilt_structural_diagram}
Structural diagram of pan-tilt rotations \\
with respect to the upper-left corner of the checkerboard.}
\end{figure}

Our goal is to model the rotational movement of a pan/tilting platform, so that we can estimate the pose of the RGB-D camera mounted on top of it. The platform rotates about arbitrarily assembled, independent axes \cite{davis2003calibrating} with loosely coupled kinematics. The structural model of such a setup is illustrated in Figure~\ref{fig:pan_tilt_structural_diagram}.  

\subsection{Rotation Parameters Acquisition}

To model the movement of the motor rotation, we first calibrate parameters for the \emph{pan} and \emph{tilt} rotation. The calibration is a two-step process, where we first estimate the direction vector of the rotation, and then estimate the center of the circular trajectory. The directions and centers of two rotations are all free variables.

When arbitrary points are rotated around a rotation axis, they create a closed circular rotation trajectory on a 3-dimensional plane, where the plane is perpendicular to the rotation axis and the circle center lies on the rotation axis.
From the coordinates of a same point in rotated frames, the circular trajectory can be obtained using the least squares method\cite{schaffrin2006note, dai2013calibration}.

During the calibration, the camera captures multiple frames of a large checkerboard in front while it rotates,  so that the checkerboard moves from one end of the field of view to another. Since the structure of the checkerboard is pre-known, all the rotation trajectories can be represented with respect to that of the top-leftmost corner. Then, we can parametrize the rotation with every corner of every frame and solve the objective function as a whole.
If the checkerboard comprises $m$ corners in the vertical direction and $n$ corners in the horizontal direction and $l$ frames were taken throughout the calibration, we have total $l\times m\times n$ corners to globally optimize. 

For the rotation of the upper-left corner, let us denote its rotation direction vector as $n = [n_x,\ n_y,\ n_z]^\intercal,\ ||n|| = 1$ and rotation circle center as $p = [a,\ b,\ c]^\intercal$.
Then the rotation axis equation becomes
\begin{equation}
\frac{x - a}{n_x} = \frac{y - b}{n_y} = \frac{z - c}{n_z},
\end{equation}
and the  rotation plane, which the upper-left corner is on, is $d$ away from the origin:
\begin{equation}
n_xx + n_yy + n_zz + d = 0. 
\end{equation}

Since all the rotation circles made from checkerboard corners are defined on the same rotation axis, the distance between their planes can be defined with respect to the indices of checkerboard corners. 
Let us denote the distances between the planes are $d_h$ and $d_w$ respectively in vertical and horizontal directions.
Then for the corner at the $i$-th row and $j$-th column of the checkerboard, the rotation circle center becomes $p_{ij} = [a_{ij},\ b_{ij},\ c_{ij}]^\intercal$, where
\begin{equation}\label{eq:rotation_circle_center}
\begin{gathered}
a_{ij} = a - n_x(i\times d_h + j\times d_w), \\
b_{ij} = b - n_y(i\times d_h + j\times d_w), \\
c_{ij} = c - n_z(i\times d_h + j\times d_w).\\
\end{gathered}
\end{equation}
The ideal rotation trajectory will be modeled as a great circle, which is represented as the intersection of a plane and a sphere in 3D space.
Here, the plane can be modeled as 
\begin{equation}
n_xx + n_yy + n_zz + d +i\times d_h + j\times d_w = 0
\end{equation}
and the sphere, or the intersecting circle can be modeled as 
\begin{equation}
(x - a_{ij})^2 + (y - b_{ij})^2 + (z - c_{ij})^2 = r_{ij}^2.
\end{equation}
Let us denote the 3D vertex of the corner at the $i$th row and $j$th column of the $k$th captured checkerboard frame as $v_{ijk} = [x_{ijk},\ y_{ijk},\ z_{ijk}]^\intercal$. 
Then, we can setup the objective function where our goal is to find parameters $n_x,\ n_y,\ n_z,\ d,\ d_h,\ d_w$ that minimize the following error for the plane model:
\begin{equation}\label{eq:rotation_plane_error}
\sum\limits_{k=0}^{l-1}\sum\limits_{i=0}^{m-1}\sum\limits_{j=0}^{n-1}(n_xx_{ijk} + n_yy_{ijk} + n_zz_{ijk} + d +i\times d_h + j\times d_w)^2.
\end{equation}
From Equation~\ref{eq:rotation_circle_center} and Equation~\ref{eq:rotation_plane_error}, one can calculate parameters $a,\ b,\ c,\ r_{ij}$ that minimize the following error for the circle model:
\begin{equation}\label{eq:rotation_circle_error}
\sum\limits_{k=0}^{l-1}\sum\limits_{i=0}^{m-1}\sum\limits_{j=0}^{n-1}((x_{ijk} - a_{ij})^2 + (y_{ijk} - b_{ij})^2 + (z_{ijk} - c_{ij})^2 - r_{ij}^2)^2.
\end{equation}
The global least squares method is adopted to minimize errors of two objective functions with regard to entire coordinate variations of 70 checkerboard corners in all frames. This yields optimized parameters for the rotation axis calibration \cite{chen2014rotation}.

\subsection{Pan-Tilt Transformation Model}

If we denote a 3D point taken in some \emph{local} camera frame after rotating \emph{tilt} and \emph{pan} angles $P_{local}$, and the point before rotations $P_{world}$, the relationship can be written with the rotation model as
\begin{equation}\label{eq:pan_tilt_model}
\begin{gathered}
\begin{bmatrix}
P_{world}  \\ 
1 
\end{bmatrix}
= T_{pan}\ R_{pan}\ T^{-1}_{pan}\ T_{tilt}\ R_{tilt}\ T^{-1}_{tilt}\ 
\begin{bmatrix}
P_{local}  \\ 
1 
\end{bmatrix}, \text{ where } \\
R(\theta) = 
\resizebox{.8\hsize}{!}{$
\setlength{\arraycolsep}{1pt}
\begin{bmatrix}
C+n_x^2(1-C) & n_xn_y(1-C)-n_zS & n_xn_z(1-C)+n_yS & 0\\
n_yn_x(1-C)+n_zS & C+n_y^2(1-C) & n_yn_z(1-C)-n_xS & 0\\
n_zn_x(1-C)-n_yS & n_zn_y(1-C)+n_xS & C+n_z^2(1-C) & 0\\
0 & 0 & 0 & 1
\end{bmatrix}
$}, \\
C = \cos\theta,\ S = \sin\theta,
\text{ and } T = 
\resizebox{.23\linewidth}{!}{$
\begin{bmatrix}
1 & 0 & 0 & a\\
0 & 1 & 0 & b\\
0 & 0 & 1 & c\\
0 & 0 & 0 & 1
\end{bmatrix}.
$}
\end{gathered}
\end{equation}

\begin{figure}[!htbp]
\centering
\begin{subfigure}[b]{0.26\linewidth}
\centering
\includegraphics[width=0.99\linewidth]{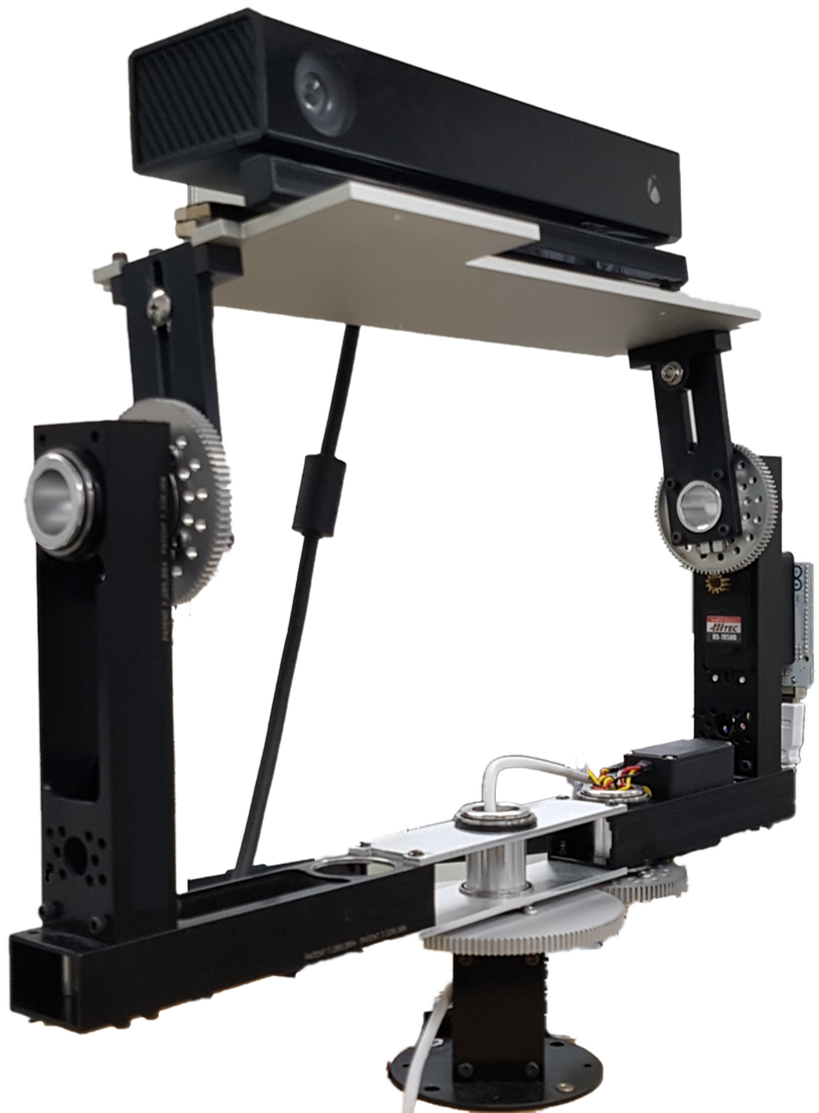}
\caption{Pan-tilt camera system.}
\end{subfigure}
~
\begin{subfigure}[b]{0.55\linewidth}
\centering
\includegraphics[width=0.99\linewidth]{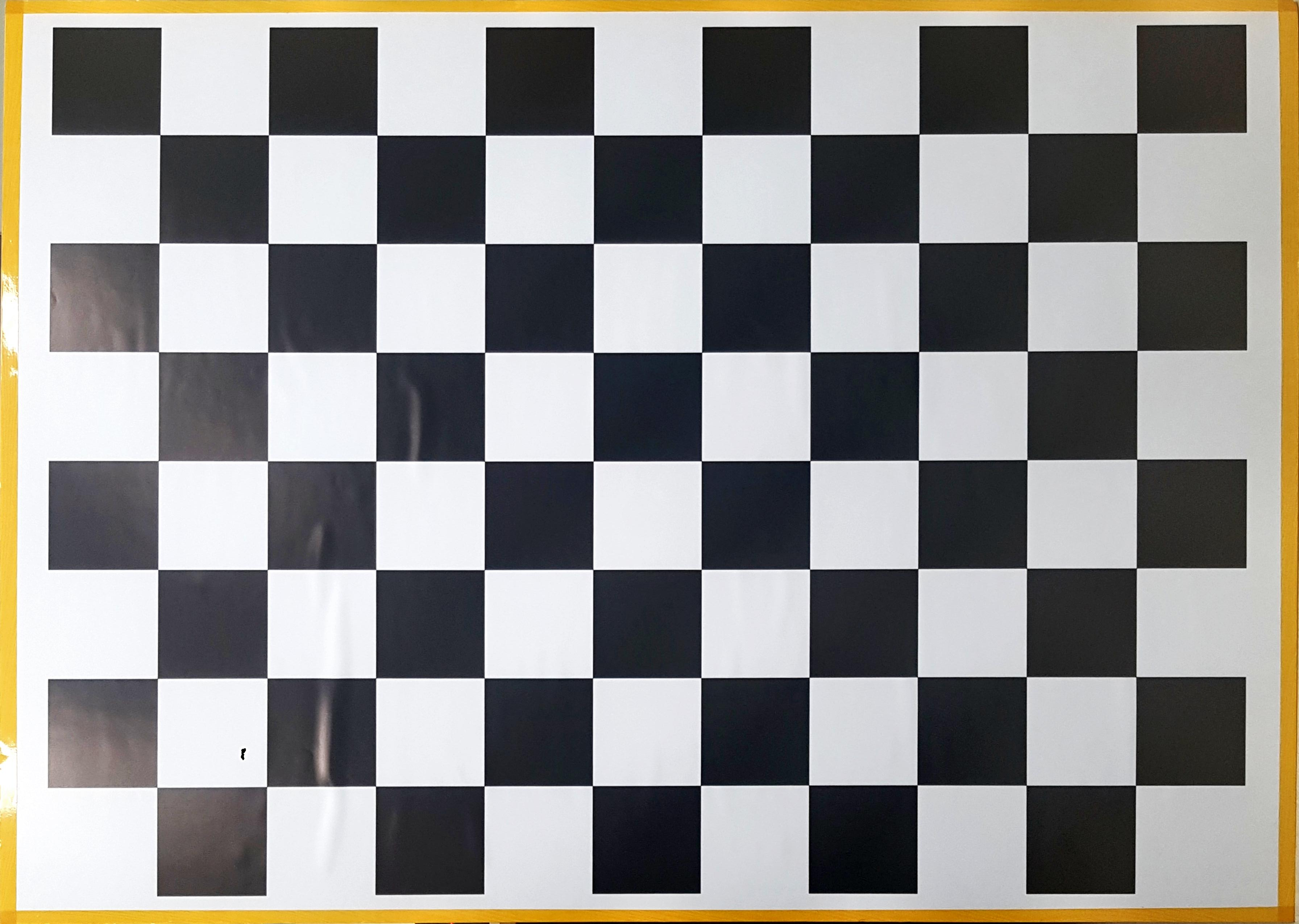}
\caption{Large checkerboard.}
\end{subfigure}

\caption{\label{fig:pantilt_experiment_configuration}
Pan-tilt experiment configuration.}
\end{figure}

Here, $R_{tilt}$ is a 4$\times$4 matrix that rotates around the direction vector $n = [n_x,\ n_y,\ n_z]^\intercal$ of the \emph{tilt} axis and $T_{tilt}$ is a 4$\times$4 matrix that translates by the coordinates of the pivot point $p = (a,\ b,\ c)$ of the \emph{tilt} axis. Transformations regarding \emph{pan} are analogous.
Note that in the model \emph{tilt} rotation comes before \emph{pan} rotation. This is due to the kinematic configuration of the pan-tilt system we used. As shown in Figure~\ref{fig:pantilt_experiment_configuration}, the tilting servo is installed on the panning arm. Thus to ensure unique rotation in \emph{world} space, we \emph{tilt} first, then \emph{pan}.

\subsection{Servo Control with Inverse Kinematics}

Our scenario is that the users want to orient the camera so that the target object is located at the image center after the rotation. Specifically, the optical axis of the camera should pass through the target point. Then the task is rotating some point on the optical axis to a target point in \emph{world} space.
This task can be thought of as rotating the linear-actuated end-effector $P_{local} = [0,\ 0,\ sz]^\intercal$ (some point on the optical axis) of a robot arm (pan-tilting platform) to the target point $P_{dst} = [dx,\ dy,\ dz]^\intercal$, with unknown $\alpha$ \emph{pan} and $\ \beta$ \emph{tilt} angles.
Using inverse kinematics notation on Equation~\ref{eq:pan_tilt_model}, the problem becomes to find the parameter vector $\theta = [\alpha,\ \beta,\ sz]^\intercal$ of 2 rotations and 1 translation that minimizes the error $\left\Vert e \right\Vert = \left\Vert P_{dst} - P_{world} \right\Vert$. We adopt the Jacobian transpose method \cite{buss2004introduction} for Algorithm~\ref{alg:inverse_kinematics}. 

\begin{algorithm}
\caption{Inverse kinematics}
\label{alg:inverse_kinematics}
\hspace*{\algorithmicindent} \textbf{input:} target point position $[dx,\ dy,\ dz]^\intercal$\\
\hspace*{\algorithmicindent} \textbf{output:} \emph{pan} angle $\alpha$, \emph{tilt} angle $\beta$, optical axis length $sz$
\begin{algorithmic}
\While{$P_{world}$ is too far from $P_{dst}$}
\State Calculate error vector $e = P_{dst} - P_{world}$.
\State Setup parameter vector $\theta = [\alpha,\ \beta,\ sz]^\intercal$
\State Compute Jacobian $J(e, \theta)$ of Equation~\ref{eq:pan_tilt_model} for the current pose
\State Pick appropriate step  $k = <e, JJ^\intercal e>/<JJ^\intercal e, JJ^\intercal e>$
\State Calculate changes in parameters, $\triangle\theta = kJ^\intercal e$
\State Apply changes to parameters $\theta\leftarrow\theta+\triangle\theta$
\State Compute new $P_{world}$ position with forward kinematics~\ref{eq:pan_tilt_model}
\EndWhile
\end{algorithmic}
\end{algorithm}

In our practice, we initialized $\alpha, \beta$ as 0 and $sz$ as $-\left\Vert P_{dst} \right\Vert$ and terminated the optimization if $\left\Vert e \right\Vert < 1\ mm$. Maximum iteration count was set to 100, though 25 on average was enough to satisfy the condition. One pitfall here is that the magnitude difference can be extensive if the value of $sz$ is set in $mm$, for $\alpha,\ \beta$ range between $-1.57$ and 1.57 radian while $sz$ may vary from 450 to 8000 $mm$. This difference causes too much fluctuations in $\alpha, \beta$ values, leading to far-from-optimal solutions. One solution is to optimize $sz$ in meter unit so that the domain difference to angles becomes minimum.

\section{Evaluation}

\subsection{System Configuration and Calibration}

To validate the proposed setup, we set up an experimental setup that comprises an Microsoft Kinect v2 and two \emph{pan} and \emph{tilt} HS-785HB servos controlled with Arduino. 
For ease of explanation, we assume internal and external parameters of the color and depth cameras are already known.
The checkerboard consists of 8 $\times$ 11 white-and-black checkers, each 100 \emph{mm} $\times$ 100 \emph{mm} in size. 

We captured 28 frames by rotating the camera end-to-end in \emph{pan} direction, and 11 frames in \emph{tilt} rotation. With coordinates of 70 extracted 3D corners, we estimated parameter values in Equation~\ref{eq:rotation_plane_error} that form a linear function of multiple planes, using the least squares method. Then, we fitted corner trajectories to a circle with parameters in Equation~\ref{eq:rotation_circle_error}. 
In Table~\ref{tb:rotation_axis_calibration}, we show values of 6 key rotation parameters that govern pan-tilt transformation of Equation~\ref{eq:pan_tilt_model}.

\begin{table}[!htbp]
\centering
\renewcommand{\arraystretch}{1}
\begin{tabular*}{\linewidth}{l@{\extracolsep{\fill}}lll}
\hline
 & \emph{Pan} rotation & \emph{Tilt} rotation \\
\hline
\hline
$n_x$ & 0.011783038 & 0.998429941 \\
$n_y$ & 0.982956803 & -0.007633507 \\
$n_z$ & -0.183458670 & -0.055492186 \\
$a$ & -82.414993286 & -412.069976807 \\
$b$ & -458.764739990 & 153.644714355 \\
$c$ & 108.336227417 & 22.413515091 \\
\hline
\end{tabular*}
\caption{\label{tb:rotation_axis_calibration}
Rotation model parameter values in \emph{mm}.}
\end{table}

\subsection{Experiment Design}
Here, we examine the performance of the proposed pan-tilt rotation model in terms of the accurate targeting capability. 
The system is tasked to adjust its attitude so that the target point is identical to the optical of center of the camera. 
We measure errors between the estimated pixels/vertices and actual captured pixels/vertices after the rotation.
The coordinates of the 70 checkerboard corners when the system is at its rest pose, \emph{i.e.} \emph{pan=tilt=0}, are used as target points, constituting 70 trials in total.

We compare our result with the result produced using the Single Point Calibration Method (SPCM) of \cite{li2015method} as the baseline. 
To steer the system, we used inverse kinematics Algorithm~\ref{alg:inverse_kinematics} for the proposed method, while for SPCM \emph{pan, tilt} values are calculated in closed form based on its geometrical model.

\subsection{Results and Discussion} \label{sec:results_and_discussion}

\begin{figure}[htb]
  \centering
  \includegraphics[width=.99 \linewidth]{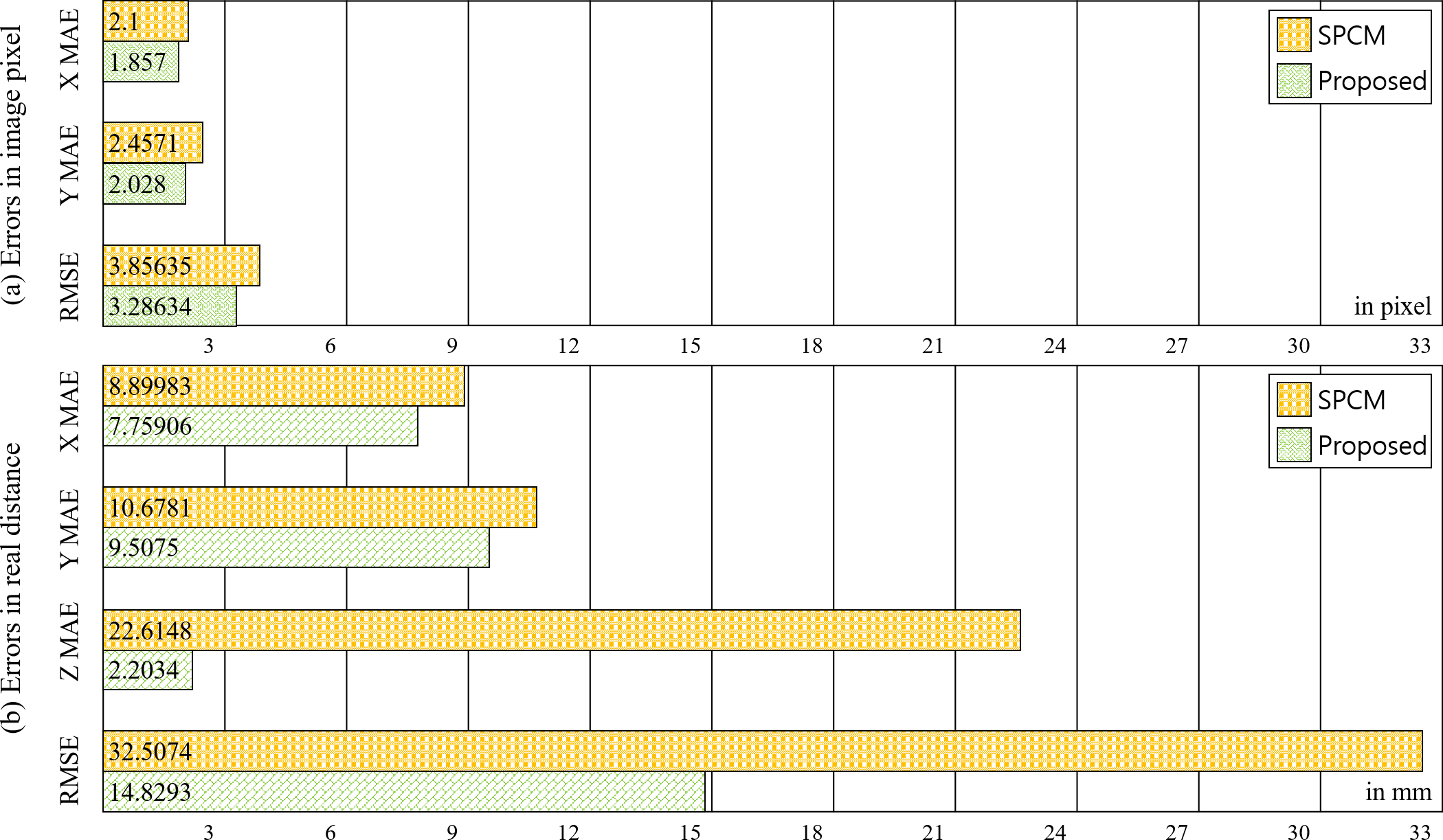}
  \caption{\label{fig:model_comparisons}
          Evaluation results of the proposed method and SPCM.}
\end{figure}

Figure~\ref{fig:model_comparisons} summarizes evaluation results of two models.
In \cite{li2015method}, authors evaluated the model with only Root Mean Squared Errors (RMSE) of L2-norms in XY image pixels. 
Since we have depth values available, we  also collected RMSEs of real distances ($mm$), and additionally measured Mean Absolute Errors (MAE) in each X, Y (and Z) directions in both pixels and $mm$s for further analysis.

In the graph, the proposed method outperforms SPCM in every metric, especially in Z directions in real distances. This can be explained by the omission of non-origin rotation centers (see Table~\ref{tb:rotation_axis_calibration}) in SPCM. Also, the rotation axes are assumed to be identical to world XY axes which lead to additional errors. The error differences become small in image planes. We explain this is due to normalizing effects of perspective projections onto the image plane.

Beside the comparative analysis, the results show seemingly high error patterns. We conjecture a number of factors have attributed to this. 
First, there is the color-depth coordinate conversion. We detected checkerboard corners in color images, then converted them into depth image coordinates. Due to limitation in Kinect SDK, conversion in sub-pixel resolution was impossible, leading to error increases. 
Second, servo rotation manipulation could be inaccurate, maybe due to unreliable kinematics of low-grade servos, such as hysteresis, jitters or inaccurate pulse width-angle mapping.

\section{Conclusion and Future Work}

In this paper, we have proposed an accurate controlling method for arbitrarily-assembled pan-tilt camera systems based on the rotation transformation model and its inverse kinematics.
The proposed model is capable of recovering rotation model parameters including positions and directions of axes of the pan-tilting platform in 3D space.
The comparative experiment demonstrates outperformance of the proposed method, in terms of accurately localizing target world points in local camera frames captured after rotations.

In following future work, we would like to extend the proposed operating mechanism to a full-scale camera pose-estimation framework that can be used in projection-based augmented reality or point cloud registration and 3D reconstruction. We would also like to delve into hardware limitations discussed in \ref{sec:results_and_discussion} and improve the pan-tilt model to compensate errors further.

\section{Acknowledgment}

This work was supported by the National Research Foundation of Korea(NRF) grant funded by the Korea government(MSIP) (No. NRF-2015R1A2A1A10055673).


\bibliographystyle{eg-alpha-doi}
\bibliography{egbibsample}


\end{document}